\documentclass[preprint,authoryear]{elsarticle}

\usepackage{amsmath,amssymb,amsfonts}
\usepackage{graphicx}
\usepackage{textcomp}
\usepackage{xcolor}
\usepackage{booktabs}
\usepackage{multirow}
\usepackage{array}
\usepackage{url}
\usepackage{listings}
\usepackage{tikz}
\usepackage{makecell}
\usepackage{tabularx}
\usetikzlibrary{shapes.geometric,arrows.meta,positioning,calc,fit,backgrounds}
\usepackage[hidelinks]{hyperref}

\graphicspath{{figures/}{./}}

\newcommand{\sysname}{MazzikaAI}

\lstdefinestyle{promptstyle}{
  basicstyle=\ttfamily\scriptsize,
  breaklines=true,
  breakatwhitespace=true,
  frame=single,
  columns=fullflexible,
  keepspaces=true,
  showstringspaces=false,
  xleftmargin=2pt,
  xrightmargin=2pt,
  aboveskip=4pt,
  belowskip=4pt,
}
\journal{Expert Systems with Applications}

\begin{document}

\begin{frontmatter}

\title{\sysname{}: A knowledge-based performance-to-prompt compiler for
real-time Arabic maqam accompaniment with a streaming text-to-music model}

\author[gvsu]{Jiaxin Du}
\ead{dujia@gvsu.edu}
\author[gvsu]{Boulbaba Abdeljaouad}
\ead{boulbeba.abdeljaouad@gmail.com}
\author[gvsu]{Yong Zhuang}
\ead{zhuangyo@gvsu.edu}
\author[gvsu]{Haoyu Li}
\ead{lihao@gvsu.edu}
\affiliation[gvsu]{organization={Grand Valley State University},
            city={Allendale},
            state={MI},
            country={USA}}

\begin{abstract}
Arabic maqam music—microtonal, modal, and built on ornamented call-and-response—is among the traditions most underserved by generative music models, whose training frameworks remain predominantly Western and equal-tempered. Real-time accompaniment sharpens this gap: an AI partner must listen, adapt dynamically, and respect idiomatic microtonal structures. Streaming text-to-music models provide strong generative capabilities but lack precise control interfaces. We present \sysname{}, a knowledge-based system that uses natural language as the actuator of a real-time control loop. By compiling live MIDI, gesture, and inferred harmony into continuously updated text prompts, \sysname{} steers an unmodified streaming generator (Google Lyria RealTime) without requiring model fine-tuning. The system embeds expert knowledge of six core maqamat, characteristic ornaments, and ensemble dynamics, maintaining real-time responsiveness with sub-second key-to-audible-update latency. Empirical evaluations demonstrate that dynamic prompt compilation reliably grounds generation in microtonal scales, significantly increasing off-grid quarter-tone content over baseline generation. Beyond its core implementation, \sysname{} illustrates how deterministic knowledge-based rules can effectively bridge expert, non-Western musical traditions and un-fine-tuned foundation models. This architecture establishes a scalable paradigm for real-time human-AI co-creation, offering a generalizable blueprint for interactive accompaniment, adaptive music education, and culturally inclusive generative audio across diverse global idioms.
\end{abstract}


\end{frontmatter}

\section{Introduction}
\label{sec:introduction}

A good accompanist listens continuously, anticipates, and supports the soloist
without overtaking them. Reproducing that behaviour computationally has
motivated decades of work on interactive music systems, from rule- and
agent-based improvisers~\citep{lewis2000voyager,rowe1993interactive} to
statistical style companions~\citep{pachet2003continuator,thom2000bob}. Neural
music models have transformed the raw \emph{quality} of what can be generated:
text-conditioned systems synthesize convincing full-band audio across a wide
stylistic range~\citep{agostinelli2023musiclm,copet2023musicgen,evans2024stableaudio}.
The property that makes them powerful---end-to-end generation of audio from a
coarse conditioning signal---also makes them hard to \emph{steer} at the
temporal grain a live performance demands. A model that renders a beautiful
minute of music from one prompt is not, by itself, a bandmate that reacts to
the note just played.

Streaming, prompt-steerable models such as Lyria RealTime~\citep{lyria2024} and
Magenta RealTime~\citep{magentart2024} narrow the gap: they emit audio
incrementally and accept prompt updates mid-session. But they supply a
\emph{control surface} without a \emph{control policy}. They answer ``given
this prompt now, what audio streams out?'' and leave open the question a live
accompaniment system must answer: \emph{how should that prompt be produced and
continuously revised from what a human is playing at this instant?} Driven by a
static prompt, a streaming model produces a generic backing track that ignores
the soloist. Symbolic and score-conditioned
models~\citep{huang2019musictransformer,roberts2018musicvae,hadjeres2017deepbach}
offer fine control but forgo the timbral realism of audio models, and adapting
either family to a new idiom normally means collecting data and fine-tuning.
The problem is sharper outside the Western tonal mainstream: Arabic maqam
music, built on microtonal intervals, characteristic ornaments, and modal
resting tones, is poorly represented in both the training distributions and the
control vocabularies of mainstream
models~\citep{shahriar2021maqam,ap2016makam}.

\paragraph{Approach} Our premise is that natural language is a
model-agnostic control surface for this class of models, and that it can be
driven as a \emph{control law} rather than authored as a query. If a system can
continuously translate what a musician is doing---register, dynamics, the
phrase just ended, the mode, the chord being held---into precise,
musically-literate prose, a general streaming model becomes a responsive band.
This reframes interactive accompaniment as \emph{compilation}: build a compiler
from live performance state to text prompts and let an off-the-shelf generator
do the synthesis. The intelligence lives in the translation layer---in knowing
that a half-flat second degree is the expressive soul of maqam bayati, that an
answering phrase should echo the soloist's last pitches, and that naming only
the active instruments prevents the model from defaulting to a full band.

Viewed from the expert-systems side, the resulting design has a familiar shape
and one unfamiliar component. \sysname{} is organized as a classical
knowledge-based system~\citep{ebcioglu1988choral,rowe1993interactive}: an
explicit, hand-authored \emph{knowledge base}; a \emph{working memory} holding
the performance state estimated from the live event stream; and a
deterministic, rule-based \emph{inference layer}. What is new is the effector.
Where a classical expert system renders its conclusions through a symbolic
synthesizer, \sysname{} renders them through a large pretrained generator
addressed in prose---a neuro-symbolic division of labour in which inspectable
symbolic expertise supplies idiom and moment-to-moment intent while the
foundation model supplies raw musical competence.

\sysname{} realizes this as a deployed system (Fig.~\ref{fig:arch}). A human
plays a MIDI keyboard in the browser; a React front end captures note events,
sustain, hand gestures, and voice commands and streams them over a WebSocket to
a Python backend. The backend maintains a rolling estimate of the
performance---density, register, motion, dynamic arc, phrase boundaries,
inferred harmony---and, on each meaningful change, compiles that state into a
prompt streamed to a text-to-music model (i.e. Lyria RealTime). Two generative streams run concurrently: a
\emph{melodic} ensemble that follows and answers the soloist, and a
tempo-locked \emph{percussion} bed. Returned audio is decoded and scheduled for
low-latency playback beneath the player's own piano. The system is designed
around vintage Arabic maqam performance---its deepest knowledge, bespoke
controller, and all output-quality experiments target that idiom---and drives
the model entirely through prompt design; the same compiler retargets to
eight further Western genres by authoring new prompt fragments alone
(Appendix~\ref{app:genres}).

\smallskip
\noindent\textbf{Contributions.}
\begin{itemize}
  \item \textbf{Prompt compilation as a real-time control law.} We formulate
        continuous natural-language prompt synthesis as the actuator of a
        knowledge-based controller (\S\ref{sec:method}): a deterministic
        compiler $C:(s_t,q_t)\mapsto p_t$ over an estimated performance state,
        a four-state call-and-response policy, and a coarse
        \emph{control-signature} gate that decides when re-steering the live
        stream is musically warranted. This turns an unmodified streaming
        text-to-music model into a responsive accompanist with no fine-tuning,
        and is independent of the particular generator behind it.
  \item \textbf{A maqam-centred musical knowledge base with prompt-level
        mechanisms that substitute for absent API control.} We contribute an
        inspectable, hand-authored knowledge base (Table~\ref{tab:kb}) whose
        core is the Arabic tradition: six maqamat with explicit quarter-tone
        spellings, characteristic degrees, ornaments, resting tones and
        negative guidance, and a takht ensemble with three role registers per
        instrument---extended by nine further Western modes and eight further
        genre configurations that demonstrate the adaptation cost
        (Appendix~\ref{app:genres}). Two mechanisms attempt control the model
        API does not expose:
        an ordered \emph{instrument-constraint} clause intended to stand in
        for stem-level mixing, and explicit maqam grounding that steers a
        Western-centric generator toward microtonal modal behaviour without
        training data. Our ablations quantify the effect of each
        (\S\ref{sec:eval-ablation}).
  \item \textbf{An instrumented evaluation with controlled ablations and
        expert playing sessions.} From end-to-end instrumentation of live
        sessions we report a full latency decomposition
        (Figs.~\ref{fig:stages}--\ref{fig:responsiveness}) and gating and
        streaming-stability measurements (Fig.~\ref{fig:gating}). A replay
        harness then drives \emph{input-identical} sessions through ablated
        variants (\S\ref{sec:eval-ablation}), measuring maqam grounding
        (significantly more quarter-tone melodic content), gating (API
        economy, with stability robust even to ungated re-steering), and the
        compiler's contribution (a static prompt collapses responsiveness).
        Playing sessions with two expert musicians provide a perceptual
        evaluation that localizes the principal remaining gap at beat-level
        entrainment (\S\ref{sec:eval-pilot}).
\end{itemize}

\noindent \S\ref{sec:related} positions the system;
\S\ref{sec:architecture} gives the architecture; \S\ref{sec:method} the
compiler; \S\ref{sec:implementation} the implementation;
\S\ref{sec:evaluation} the measurements, ablation experiments, and
perceptual evaluation; \S\ref{sec:discussion} discussion and limitations;
Appendix~\ref{app:genres} catalogues the genre configurations.

\section{Related Work and Positioning}
\label{sec:related}

Table~\ref{tab:related} places \sysname{} against the five families of prior
work it draws on, along the axes that matter for live accompaniment: what the
system is steered \emph{through}, at what \emph{granularity}, what it costs to
\emph{adapt} to a new idiom, whether it generates \emph{streaming audio}, and
whether it supports microtonal non-Western modes.

\begin{table}[tbp] \centering \caption{Positioning of \sysname{} against related families. ``Granularity'' is the finest interval at which the system's behavior can be redirected; ``adaptation cost'' is what a new idiom requires.} \label{tab:related} \footnotesize \setlength{\tabcolsep}{3pt} \renewcommand{\arraystretch}{1.15} \begin{tabularx}{\columnwidth}{ @{} >{\raggedright\arraybackslash}p{0.18\columnwidth} >{\raggedright\arraybackslash}X >{\raggedright\arraybackslash}p{0.12\columnwidth} >{\raggedright\arraybackslash}p{0.14\columnwidth} cc @{} } \toprule \makecell[c]{\textbf{Family}} & \makecell[c]{\textbf{Control}\\\textbf{surface}} & \makecell[c]{\textbf{Granu\-larity}} & \makecell[c]{\textbf{Adaptation}\\\textbf{cost}} & \makecell[c]{\textbf{Streaming}\\\textbf{audio}} & \makecell[c]{\textbf{Micro\-tonal}} \\ \midrule Knowledge-based symbolic\textsuperscript{1} & hand-authored rules & note/voice & new rules & no & possible \\ Interactive symbolic improvisers\textsuperscript{2} & listening $+$ bespoke composer & note/beat & redesign or retrain & no & rare \\ Symbolic neural\textsuperscript{3} & latents / partial scores & note & fine-tune & no & no \\ Text-conditioned audio\textsuperscript{4} & one-shot text prompt & whole clip & fine-tune / prompt & no & no \\ Streaming audio models\textsuperscript{5} & live text $+$ scalars & chunk ($\sim$2\,s) & manual prompting & yes & untested \\ \midrule \textbf{\sysname{}} & \textbf{compiled text, closed loop} & \textbf{change-gated chunk} & \textbf{new prompt fragment} & \textbf{yes} & \textbf{grounded} \\ \bottomrule \end{tabularx} \vspace{2pt} \begin{minipage}{\columnwidth} \scriptsize \textsuperscript{1}\citep{ebcioglu1988choral,cope1992emi}; \textsuperscript{2}\citep{rowe1993interactive,lewis2000voyager, pachet2003continuator,thom2000bob,scarlatos2025realjam}; \textsuperscript{3}\citep{huang2019musictransformer,roberts2018musicvae, hadjeres2017deepbach}; \textsuperscript{4}\citep{agostinelli2023musiclm,copet2023musicgen, evans2024stableaudio}; \textsuperscript{5}\citep{lyria2024,magentart2024}. \end{minipage} \end{table}

\paragraph{Knowledge-based systems for music} Encoding musical expertise as an
explicit rule base is among the oldest ambitions of intelligent systems.
Ebcio\u{g}lu's CHORAL~\citep{ebcioglu1988choral} harmonized Bach chorales with
hundreds of hand-authored rules; Cope's EMI~\citep{cope1992emi} recombined
corpus signatures under rule-based control; Rowe~\citep{rowe1993interactive}
organized machine musicianship as listening, interpretation, and rule-governed
response. \sysname{} deliberately retains the two properties these share---the
knowledge is \emph{inspectable} and the inference is \emph{deterministic}---and
replaces only what historically bounded them, the symbolic effector, with a
pretrained streaming generator addressed in prose, in the spirit of recent
efforts to unify foundation models with explicit knowledge
representations~\citep{pan2024llmkg}.

\paragraph{Interactive and co-creative systems} Voyager~\citep{lewis2000voyager}
negotiated musical agency alongside a human; the Continuator~\citep{pachet2003continuator}
learned a player's style online and extended their phrases;
BoB~\citep{thom2000bob} traded solos with a musician; and
ReaLJam~\citep{scarlatos2025realjam} recently demonstrated live human--AI
jamming with a reinforcement-learning-tuned transformer. The turn-taking
dynamics these systems exploit are well documented
ethnomusicologically~\citep{monson1996saying}. \sysname{} performs the same
listen-then-respond cycle, but couples the listening machinery to a large
pretrained \emph{audio} model mediated by language rather than to a bespoke
symbolic composer.

\paragraph{Neural music generation and streaming} Symbolic models predict note
events; raw-audio and latent-audio models---WaveNet~\citep{oord2016wavenet},
Jukebox~\citep{dhariwal2020jukebox}, AudioLM~\citep{borsos2023audiolm}, and
the text-conditioned MusicLM~\citep{agostinelli2023musiclm},
MusicGen~\citep{copet2023musicgen}, Stable Audio~\citep{evans2024stableaudio},
Riffusion~\citep{forsgren2022riffusion}, built over neural codecs such as
EnCodec~\citep{defossez2022encodec}---model timbre and production directly but
are designed for offline, one-shot synthesis. Streaming
models~\citep{lyria2024,magentart2024} are the closest substrate; \sysname{}
is built on Lyria RealTime. The Live Music Models
report~\citep{magentart2024} describes two adjacent capabilities: audio
injection, which conditions the model on raw input audio rather than on an
interpreted model of the performance, and manual prompt steering by a human
operator. Closing that loop \emph{automatically and symbolically}---estimating a
structured performance state and compiling it, through an explicit knowledge
base, into a continuously refreshed prompt and generation config---is the gap
\sysname{} fills.

\paragraph{Prompting as control} Prompting emerged in NLP as a way to steer
frozen models~\citep{brown2020gpt3,liu2023promptsurvey}, but almost always as a
\emph{one-shot, text-in--text-out} operation. \sysname{} extends it along two
axes at once: from one-shot authoring to a \emph{continuous closed-loop} regime
in which the prompt string is the actuator of a real-time feedback law, and
from text output to streaming \emph{audio}. To our knowledge, treating prompt
engineering as the control law of an interactive audio system is a framing
prior prompting work does not develop.

\paragraph{Non-Western modal music} Arabic maqam and Turkish makam are
organized around modal scales with \emph{microtonal} degrees, signature
ornaments, and resting tones. Computational work has focused largely on
analysis---the review of computational makam
research~\citep{ap2016makam} documents the tuning and intonation problems that
defeat twelve-tone tools, and recent work classifies maqamat from
audio~\citep{shahriar2021maqam}---while generation of authentic microtonal
modal music remains underexplored. \sysname{} addresses this asymmetry by
\emph{grounding} rather than retraining: prompts spell each maqam out
explicitly, naming the microtonal degrees, ornaments and resting tones and
supplying negative guidance against Western tonality
(\S\ref{sec:kb}).

\section{System Architecture}
\label{sec:architecture}

\begin{figure*}[tb]
\centering
\resizebox{\linewidth}{!}{%

\begin{tikzpicture}[
    font=\small,
    >=Latex,
    node distance=6mm,
    box/.style={
        draw,
        rounded corners=2pt,
        align=center,
        minimum height=7mm,
        inner sep=3pt,
        fill=white
    },
    io/.style={
        box,
        fill=black!5,        
    minimum width=15mm,
    minimum height=7mm,
    align=center
    },
    srv/.style={
        box,
        fill=blue!6
    },
    ext/.style={
        box,
        fill=orange!12,
        minimum width=30mm
    },
    net/.style={
        box,
        fill=green!8,
        minimum height=9mm
    },
    lbl/.style={
        midway,
        font=\scriptsize\itshape,
        fill=white,
        inner sep=1pt
    }
]

\node[io] (midi) {MIDI};
\node[io, below=of midi] (gesture) {Gesture};
\node[io, below=of gesture] (voice) {Voice};
\node[io, below=of voice] (sheet) {Sheet};

\node[net, right=14mm of midi] (ws)
    {Web Socket};

\node[io, below=31mm of ws, minimum width=32mm] (audio)
    {Web Audio out \\(dual contexts, mix)};

\node[srv, right=18mm of ws] (est)
    {PlayerSession\\(working memory)};

\node[srv, right=of est] (ctrl)
    {Controller\\$q_t,\theta_t$};

\node[srv, below=9mm of ctrl] (comp)
    {Prompt\\compiler $C$};

\node[srv, below left=11mm and 4mm of comp] (mel)
    {LyriaEngine\\(melodic)};

\node[srv, below right=11mm and 4mm of comp] (perc)
    {LyriaEngine\\(percussion)};

\node[ext, below=24mm of sheet, xshift=23mm] (omr)
    {Gemini OMR};

\node[ext, right=50mm of omr] (lyria)
    {Lyria RealTime};

\begin{scope}[on background layer]

\node[
    draw,
    dashed,
    rounded corners,
    fit=(midi)(gesture)(voice)(sheet)(ws)(audio),
    inner sep=10pt,
    label=above:{
        \footnotesize\textbf{Browser client}
    }
] (browserbox) {};

\node[
    draw,
    dashed,
    rounded corners,
    fit=(est)(ctrl)(comp)(mel)(perc),
    inner sep=9pt,
    label=above:{
        \footnotesize\textbf{FastAPI backend}
    }
] (backendbox) {};

\node[
    draw,
    dashed,
    rounded corners,
    fit=(omr)(lyria),
    inner xsep=16mm,
    inner ysep=5mm,
    label=above:{
        \footnotesize\textbf{Cloud models}
    }
] (cloudbox) {};

\end{scope}

\draw[->] (midi.east) -- (ws.west);

\draw[->]
    (gesture.east)
    -| ([xshift=-6mm]ws.west)
    -- (ws.west);

\draw[->]
    (voice.east)
    -| ([xshift=-8mm]ws.west)
    -- (ws.west);

\draw[->]
    (sheet.east)
    -| ([xshift=-10mm]ws.west)
    -- (ws.west);

\draw[->]
    (ws.east) -- (est.west)
    node[lbl, above] {JSON};

\draw[->] (est.east) -- (ctrl.west);
\draw[->] (ctrl.south) -- (comp.north);

\draw[->]
    (comp.south west) -- (mel.north)
    node[lbl, left] {prompt};

\draw[->]
    (comp.south east) -- (perc.north)
    node[lbl, right] {prompt};

\draw[->]
    (sheet.south)
    to[out=-90, in=90]
    (omr.north);

\draw[<->]
    (mel.south)
    to[out=-90, in=180, looseness=1.15]
    (lyria.west);

\draw[<->]
    (perc.south)
    to[out=-90, in=0, looseness=1.15]
    (lyria.east);

\draw[->]
    (mel.west)
    to[out=-180, in=0, looseness=1.1]
    (audio.east);

\draw[->]
    (perc.south west)
    to[out=-135, in=-10, looseness=1.2]
    (audio.south east);

\draw[->]
    (ws.south) -- (audio.north)
    node[lbl, right] {audio\_chunk};

\end{tikzpicture}
}
\caption{End-to-end architecture. Performer input is serialized as JSON
WebSocket messages to a FastAPI backend, which updates the working memory,
selects an accompaniment state $q_t$ and generation parameters $\theta_t$, and
compiles a prompt $p_t$. Two independent Lyria RealTime sessions---a following
melodic engine and an independent tempo-locked percussion engine---return
$48$\,kHz stereo PCM that is returned as \texttt{audio\_chunk} messages and
mixed client-side beneath the performer's local piano.}
\label{fig:arch}
\end{figure*}
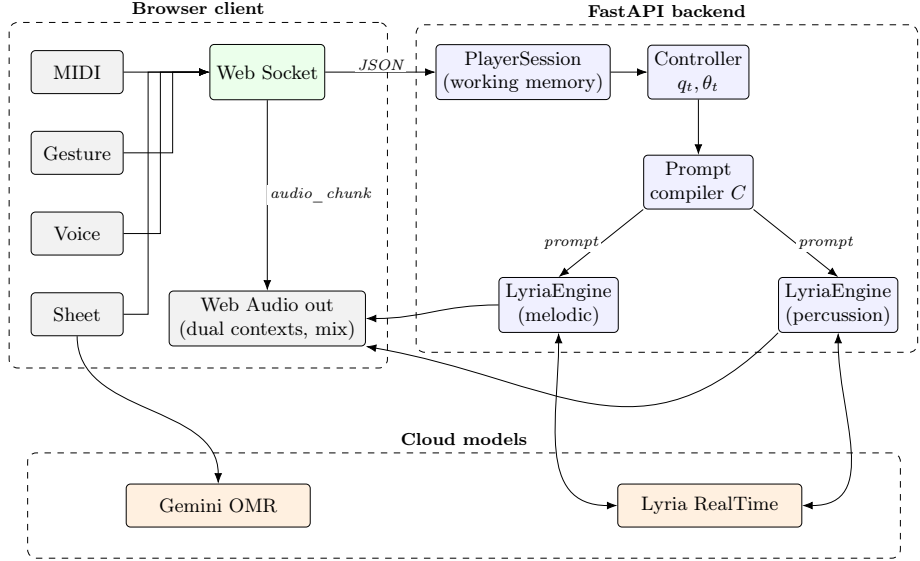

\sysname{} is a three-tier system---browser client, Python backend, cloud
generative model---connected by a single bidirectional WebSocket
(Fig.~\ref{fig:arch}). Four design commitments shape it. \emph{(i) Text is the
sole point of control}: the generator is a black box that consumes prose and
emits audio, so the deployed model can be replaced without touching the control
layer. \emph{(ii) Idiom is injected, not learned}: every prompt carries explicit
grounding for the active mode, including negative guidance against Western
tonality. \emph{(iii) The performer conducts}: rather than a fixed band, the
player toggles ensemble sections in and out and the compiler actively suppresses
the rest. \emph{(iv) Re-steering is gated}: a coarse control signature decides
when the live stream may be re-prompted, trading responsiveness against
stability.

\paragraph{Client and transport} A React/TypeScript application captures four
input modalities: note events (\texttt{note\_on}/\texttt{note\_off}, velocity,
CC64 sustain) through the Web MIDI API~\citep{webmidi}; hand gestures
recognized on-device with MediaPipe Hands~\citep{mediapipe2019}; spoken
commands through the Web Speech API; and a Learn Mode that parses MusicXML
in-browser and delegates scanned PDF pages to Gemini optical music recognition.
The performer's own notes are rendered locally by a sampled grand piano, so the
human's sound is never delayed by the network round trip. Every input becomes a
small JSON message: \texttt{midi\_event}, \texttt{control\_update} (genre,
maqam, tempo; debounced to ${\sim}400$\,ms), \texttt{gesture},
\texttt{section\_toggle}, and a \texttt{tick} heartbeat every $250$\,ms that
advances time-dependent state when no notes arrive. The backend replies with
\texttt{state\_update} messages (estimated state and compiled prompt) and
\texttt{audio\_chunk} messages tagged
\texttt{source}$\in$\{\texttt{melody},\texttt{percussion}\}.

\paragraph{Backend and output} A FastAPI server maintains a
\texttt{PlayerSession} whose working memory is a rolling window of the last
$128$ note events (\S\ref{sec:state}). A controller maps that state to an
accompaniment behaviour and a generation-parameter vector, and the compiler
synthesizes the prompt (\S\ref{sec:synthesis}). Prompt and parameters are pushed
to a live Lyria session whose background receive loop pulls $48$\,kHz $16$-bit
stereo PCM~\citep{lyria2024,defossez2022encodec}, WAV-packs and base64-encodes
it, and enqueues it on a bounded queue (\texttt{maxsize}$=8$, oldest dropped
when full) so buffered audio cannot grow without bound. The client decodes each
chunk with the Web Audio API and schedules playback on two separate
\texttt{AudioContext}s using a running ``next start'' cursor with a small
negative overlap ($-40$\,ms melody, $-20$\,ms percussion) intended to crossfade
chunk boundaries. The two AI streams are attenuated ($0.45\times$ melody,
$0.38\times$ percussion) beneath the performer's local piano, so arbitration
between the voices is a deterministic gain decision rather than a generative
one.

\subsection{The Accompaniment State Machine}
\label{sec:statemachine}

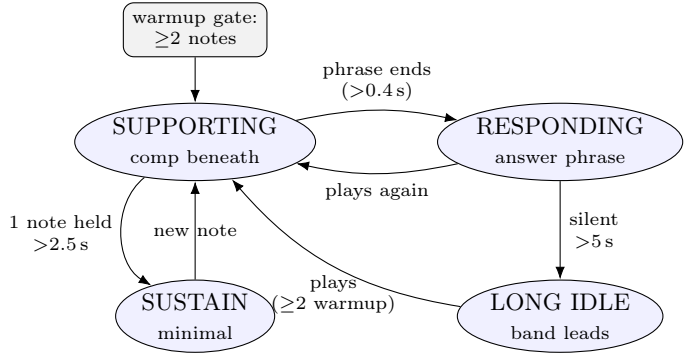
\begin{figure}[tb]
\centering
\begin{tikzpicture}[
  font=\small, >=Latex, node distance=13mm and 16mm,
  st/.style={draw, ellipse, align=center, minimum width=18mm,
             minimum height=9mm, inner sep=1pt, fill=blue!6},
  every edge/.style={draw, ->},
  edgelbl/.style={font=\scriptsize, align=center},
]
\node[st] (supp)  {SUPPORTING\\{\scriptsize comp beneath}};
\node[st] (resp)  [right=of supp] {RESPONDING\\{\scriptsize answer phrase}};
\node[st] (sust)  [below=of supp] {SUSTAIN\\{\scriptsize minimal}};
\node[st] (idle)  [below=of resp] {LONG IDLE\\{\scriptsize band leads}};
\draw (supp) edge[bend left=12] node[edgelbl,above]{phrase ends\\($>$$0.4$\,s)} (resp);
\draw (resp) edge[bend left=12] node[edgelbl,below]{plays again} (supp);
\draw (supp) edge[bend right=55] node[edgelbl,left]{1 note held\\$>$$2.5$\,s} (sust);

\draw (sust) edge node[edgelbl]{new note} (supp);
\draw (resp) edge node[edgelbl,right]{silent\\$>$$5$\,s} (idle);
\draw (idle) edge[bend left=20] node[edgelbl,below]{plays\\($\geq$$2$ warmup)} (supp);
\node[draw, rounded corners, fill=black!5, font=\scriptsize, align=center]
      (start) [above=6mm of supp] {warmup gate:\\$\geq$$2$ notes};
\draw (start) edge (supp);
\node[font=\scriptsize\itshape, align=center] (stop) [below=5mm of idle]
      {melodic stream stops after $>$$4$\,s silence};
\end{tikzpicture}
\caption{The four-state accompaniment policy $\pi$. Transitions are driven by
playing and silence timers over the performance stream; a warmup gate admits
the ensemble and a silence stop halts the melodic stream until the soloist
resumes.}
\label{fig:statemachine}
\end{figure}

The controller resolves the estimated performance state into one of four
behaviours (Fig.~\ref{fig:statemachine}). \textsc{Supporting}: while the
soloist plays, the band comps quietly beneath the line at the highest
prompt-adherence and lowest density. \textsc{Responding}: when a phrase ends, an
answer window opens ${\sim}0.3$--$2.5$\,s afterward, in which the prompt lists
the soloist's just-played pitches and instructs an expressive echo.
\textsc{Sustain}: a single note held for ${>}2.5$\,s is read as a held tone and
the accompaniment stays minimal. \textsc{LongIdle}: after ${>}5$\,s of silence
with sections active, the band assumes the floor and leads freely at the
highest density. Two gates bound the machine: a \emph{warmup gate} requires two
notes after a section is conducted in before the ensemble may sound, and a
\emph{silence stop} halts the melodic stream after ${>}4$\,s of silence,
forcing a fresh re-prompt when the player returns.

\subsection{Why Two Decoupled Streams}
\label{sec:concurrency}

\sysname{} runs two independent Lyria sessions because melodic accompaniment
and rhythmic bed obey different clocks. The \emph{melodic} engine has
\texttt{mute\_drums} enabled and no fixed BPM lock; it is re-prompted as
harmony, register and phrasing evolve. The \emph{percussion} engine, once
triggered by gesture, runs continuously as a tempo-locked $4/4$ bed at the
selected BPM, independent of the soloist, cycling through escalating style
patterns and using tighter generation parameters to stay literal and sparse.
Their control signals genuinely conflict: the melodic voice must be free to
breathe, slow, or rest to follow the performer, whereas a groove must remain
metronomically steady, and folding both into one prompt would force one to
compromise the other. Separation also isolates a hard model constraint---Lyria
cannot change BPM on a live stream, so a tempo change forces a full
disconnect-and-reconnect, which the percussion engine can absorb without
interrupting the melodic conversation.

\section{Method: Compiling Performance into Prompts}
\label{sec:method}

\subsection{Problem Formulation}
\label{sec:problem}

We treat companion accompaniment as a real-time transduction problem. Let the
soloist produce timestamped events $e_1,e_2,\dots$, each a MIDI note-on/off with
velocity, a sustain-pedal change, a gesture, or a control update. Rather than
condition the generator on the raw stream, the system maintains an estimated
performance state
\[
  s_t = \Phi\!\left(\{e_i : t - W \le \tau_i \le t\}\right),
\]
where the estimator $\Phi$ (\S\ref{sec:state}) runs over a bounded rolling
window---a deque of the last $128$ note events, with feature-specific look-back
horizons $W$ of $2$--$30$\,s. A discrete controller $\pi$ maps the state to one
of four accompaniment behaviours,
$q_t = \pi(s_t) \in \{\textsc{LongIdle},\textsc{Responding},\textsc{Supporting},
\textsc{Sustain}\}$, and $\Theta$ maps $(q_t,s_t)$ to a genre-conditioned
generation-parameter vector
$\theta_t = (\text{temperature},\text{top-}k,\text{guidance},\text{density},
\text{brightness})$ (\S\ref{sec:controller}).

The method's core is a deterministic \emph{prompt compiler}
\[
  C : (s_t, q_t) \;\longmapsto\; p_t \in \Sigma^{*},
\]
a pure, side-effect-free function emitting a single natural-language string over
the alphabet $\Sigma$ of English text (\S\ref{sec:synthesis}). Determinism is
load-bearing: identical states yield identical prompts, which is what makes the
downstream gate sound. The generator $G$---invoked as a streaming,
weighted-prompt latent music model~\citep{lyria2024}---produces
$a_t = G(p_t,\theta_t)$. Two instances of $G$ run concurrently
(\S\ref{sec:concurrency}). To avoid destabilizing the live stream, compiler
output reaches $G$ only when a coarse \emph{control signature} changes:
\[
  \text{re-prompt at } t \iff
  \sigma(s_t,q_t,\theta_t) \neq \sigma(s_{t^-},q_{t^-},\theta_{t^-}).
\]
State estimation, discrete control, deterministic text synthesis, gated
streaming: these four stages structure the rest of the section.

\subsection{Working Memory: Performance-State Estimation}
\label{sec:state}

Estimation is driven by two inputs: MIDI events, which mutate state on arrival,
and the $250$\,ms heartbeat, which recomputes time-dependent features when no
note arrives. Table~\ref{tab:state} lists the principal features and their
horizons.

Register is the mean active pitch (\texttt{low}${<}50$, \texttt{high}${>}68$,
else \texttt{mid}) and texture the number of simultaneously held keys. Over
short windows the estimator derives motion (sign of summed pitch differences of
the last five onsets), melodic centre and its direction, playing speed from
notes-per-second over $3$\,s, and a dynamic arc by comparing the early and late
halves of the last twelve velocities. Phrase segmentation is silence-driven:
inter-onset gaps below $0.35$\,s are legato continuation, gaps above $0.4$\,s
while silent mark a phrase end, and gaps above $2.5$\,s mark idle. A phrase end
more than $1.2$\,s after the previous one increments a verse counter that
quantizes into a \texttt{verse\_stage} (\texttt{opening}/\texttt{developing}/%
\texttt{full}), so the accompaniment opens up as the session matures. At each
counted phrase end the system captures \texttt{last\_phrase\_pitches}, the
unique pitches of the just-ended phrase in order, so the ensemble can echo them
literally.

For harmonically grounded genres the estimator performs two independent forms of
chord inference. \emph{Real-time recognition} matches the pitch-class set of
currently held keys against interval templates, trying every pitch class as
candidate root and scoring higher-information chords first (dominant/minor
seventh $>$ triad $>$ power chord $>$ dyad $>$ octave), yielding a symbol such
as \texttt{A7} or \texttt{Dm}. When no chord is held, \emph{implied-chord
inference} estimates the current region of the twelve-bar blues form by scoring
the recent pitch-class histogram against the three functional harmonies, with
each tonic weighted $\times3$:
\[
\begin{aligned}
\text{score}_{\text{A7 (I)}} &= 3\,\#(A) + \#(E) + \#(G),\\
\text{score}_{\text{D7 (IV)}} &= 3\,\#(D) + \#(A) + \#(C),\\
\text{score}_{\text{E7 (V)}} &= 3\,\#(E) + \#(B) + \#(D),
\end{aligned}
\]
returning the arg-max region.

Two derived features exist purely to stabilize the downstream stream. The
\emph{harmony fingerprint} is a coarse pitch-class snapshot over the last
$2$\,s, quantized into $2$-semitone buckets---%
\texttt{tuple(sorted(set((p\,\%\,12)//2 for p in pitches)))}---so that adjacent
semitones collapse and the fingerprint is meant to change only when harmony
genuinely shifts. It replaced an earlier key that changed on every note-on and
re-prompted the model continuously during melodic runs; \S\ref{sec:eval-gating}
reports how far short of its intent it falls. The \emph{style profile} is a
natural-language sentence recomputed every two verses from $30$\,s of history
using genre-aware vocabulary, e.g.\ \texttt{"ascending blues runs climbing
toward the high note, wide bends and leaps, wailing in the high register"}.

\begin{table}[tbp]
\centering
\caption{Working memory: principal features of the estimated performance state
$s_t$, with the horizon each is computed over.}
\label{tab:state}
\footnotesize
\begin{tabular}{@{}p{0.29\columnwidth}p{0.17\columnwidth}p{0.44\columnwidth}@{}}
\toprule
\textbf{Feature} & \textbf{Horizon} & \textbf{Values / role} \\
\midrule
\multicolumn{3}{@{}l}{\emph{Texture and dynamics}}\\
register & held keys & low / mid / high (mean pitch) \\
texture\_hint & held keys & solo / duo / small / full chord \\
motion & 5 onsets & ascending / descending / static \\
melodic\_center & 6 onsets & mean pitch (register target) \\
register\_follow & 6 onsets & rise / fall / hold \\
playing\_speed & 3\,s & fast / moderate / slow (notes\,s$^{-1}$) \\
dynamic\_arc & 12 velocities & crescendo / diminuendo / steady \\
\midrule
\multicolumn{3}{@{}l}{\emph{Phrase and session time}}\\
phrase\_state & silence gap & start / continue / end / idle \\
verse\_count $\to$ stage & session & opening / developing / full \\
call\_response\_active & 0.3--2.5\,s & post-phrase answer window \\
long\_idle & ${>}5$\,s & player silent; band should lead \\
is\_sustained\_hold & ${>}2.5$\,s & one note held \\
last\_phrase\_pitches & phrase end & pitches to echo back \\
\midrule
\multicolumn{3}{@{}l}{\emph{Harmony}}\\
detected\_chord & held keys & template-matched chord symbol \\
implied\_chord & 2\,s & 12-bar region (A7 / D7 / E7) \\
recent\_pitch\_context & buffer & last 10 unique note names \\
harmony\_fingerprint & 2\,s & coarse PC buckets; gate input \\
style\_profile & 30\,s & NL sentence, recomputed / 2 verses \\
\bottomrule
\end{tabular}
\end{table}

\subsection{Controller and Generation Parameters}
\label{sec:controller}

The controller selects $q_t$ by a simple interpretable priority
(\textsc{LongIdle} $>$ \textsc{Responding} $>$ \textsc{Supporting}, with
\textsc{Sustain} as the special case of a single held note): the player actively
playing yields \textsc{Supporting}; the open answer window or an ending mode
yields \textsc{Responding}; sustained silence with sections active yields
\textsc{LongIdle}.

Each state maps to a numeric control vector read from a per-genre table, which
stores for each of the nine genres a
$(\text{temperature},\text{top-}k,\text{guidance},\text{density})$ quadruple per
state plus three register-dependent brightness values---$135$ hand-tuned values
in total. Table~\ref{tab:params} shows a representative slice. The rationale is
consistent across genres: higher temperature is more exploratory; lower
top-$k$ and higher guidance bind the model more tightly to the prompt; density
targets note sparsity. \textsc{LongIdle} therefore uses the highest temperature
and density (the band improvises freely) while \textsc{Supporting} uses the
lowest temperature and density and the \emph{highest} guidance (the band stays
literal and out of the way). A verse-stage boost of $0.00/0.03/0.06$ is added to
density (capped at $0.90$) so the accompaniment enriches gradually, and
brightness is selected by the soloist's register so a lead voice cuts through in
the high register. The percussion engine uses markedly tighter settings (blues:
temperature $0.32$, top-$k$ $8$, guidance $10.5$, density $0.32$) so it plays
the pattern rather than improvising fills.

\begin{table}[tbp]
\centering
\caption{Representative generation parameters $\theta_t = \Theta(q_t,s_t)$, four
of nine genres. Density excludes the verse-stage boost; brightness is
register-dependent and omitted. The remaining knobs the API
exposes---\texttt{bpm} (fixed at connect) and \texttt{mute\_drums} (set for the
melodic engine)---are constant per engine.}
\label{tab:params}
\footnotesize
\begin{tabular}{@{}llcccc@{}}
\toprule
\textbf{Genre} & \textbf{State} & \textbf{temp} & \textbf{top-}$k$ & \textbf{guid.} & \textbf{dens.} \\
\midrule
\multirow{3}{*}{blues}
 & LongIdle    & 0.78 & 22 & 5.5 & 0.58 \\
 & Responding  & 0.72 & 18 & 6.0 & 0.42 \\
 & Supporting  & 0.60 & 12 & 7.0 & 0.22 \\
\midrule
\multirow{3}{*}{vintage arabic}
 & LongIdle    & 0.72 & 20 & 5.8 & 0.48 \\
 & Responding  & 0.65 & 16 & 6.5 & 0.38 \\
 & Supporting  & 0.52 & 12 & 7.5 & 0.20 \\
\midrule
\multirow{3}{*}{jazz fusion}
 & LongIdle    & 0.82 & 24 & 5.0 & 0.55 \\
 & Responding  & 0.74 & 20 & 5.8 & 0.40 \\
 & Supporting  & 0.64 & 14 & 6.8 & 0.22 \\
\midrule
\multirow{3}{*}{classical}
 & LongIdle    & 0.60 & 16 & 6.5 & 0.45 \\
 & Responding  & 0.55 & 14 & 7.0 & 0.32 \\
 & Supporting  & 0.48 & 10 & 8.0 & 0.18 \\
\bottomrule
\end{tabular}
\end{table}

\paragraph{Re-prompt gating} A live stream cannot tolerate re-prompting on every
note. The controller therefore emits a coarse control signature $\sigma$---a
tuple of only the musically meaningful determinants of the output: genre, maqam,
the \textsc{LongIdle}/\textsc{Responding}/answer-window flags, the harmony
\emph{fingerprint} rather than raw notes, the first four echo pitches, the
active section set, verse stage, density and brightness rounded into coarse
buckets, BPM, and the detected and implied chords. The transport layer pushes a
new prompt only when $\sigma$ changes. By design this makes the signature
invariant to the note-by-note churn of a melodic run and sensitive only to
genuine change; \S\ref{sec:eval-gating} measures how well that holds.

\subsection{The Knowledge Base}
\label{sec:kb}

Everything the system knows about music is hand-authored prose and numbers,
inventoried in Table~\ref{tab:kb}. Its core is the Arabic tradition: six
maqamat with quarter-tone spellings and a five-instrument takht
configuration---cello, oud, strings, qanun, and nay (lead)---driven by a
bespoke controller with idiom-specific harmonic and echo logic. Around that
core the same structures carry eight further Western genres ($15$ mode
descriptions, $38$ ensemble sections with three role registers each, $135$
generation parameters, and $36$ percussion patterns in total); the full genre
catalogue, which exists chiefly as evidence that retargeting costs only new
prompt fragments, is given in Appendix~\ref{app:genres}. The knowledge base
is small enough to read in an afternoon.

\begin{table}[tbp]
\centering
\caption{Inventory of the hand-authored knowledge base. Every entry is
human-readable text or a tuned constant; none is learned. Line counts are of the
deployed implementation.}
\label{tab:kb}
\footnotesize
\setlength{\tabcolsep}{4pt}
\begin{tabular}{@{}p{0.26\columnwidth}rp{0.50\columnwidth}@{}}
\toprule
\textbf{Component} & \textbf{Size} & \textbf{Content and prompt slot} \\
\midrule
Mode / scale descriptions & 15 &
  6 Arabic maqamat (bayati, rast, hijaz, nahawand, saba, kurd) with
  quarter-tone spellings $+$ 9 Western scales and modes; each gives note
  spelling, characteristic degree, ornaments, resting tones, colour, and
  negative guidance $\to$ \emph{maqam line} \\
Ensemble sections & 38 &
  across 9 genres; each with a label and three role registers
  (\texttt{supporting}, \texttt{responding}, \texttt{solo\_responding}),
  i.e.\ 114 role descriptions $\to$ \emph{role assignment} \\
Genre configurations & 9 &
  section order (low$\to$high), lead-preference list, header text,
  no-percussion clause (Appendix~\ref{app:genres}, Table~\ref{tab:genres}) \\
Generation parameters & 135 &
  $9$ genres $\times$ ($3$ states $\times$ $4$ knobs $+$ $3$ brightness
  values) $\to$ $\theta_t$ (Table~\ref{tab:params}) \\
Percussion patterns & 36 &
  $9$ genre families $\times$ $4$ escalating styles, each with a tempo anchor
  and a no-melody exclusion clause \\
Harmonic rules & --- &
  interval-template chord recognition, 12-bar region inference, four-tier
  blues harmonic-lock priority \\
\midrule
Implementation & \multicolumn{2}{l}{$2{,}434$ lines of compiler,
  $246$ of controller, $571$ of state estimation (Python)} \\
\bottomrule
\end{tabular}
\end{table}

\paragraph{Maqam grounding} To hold a general model inside a modal world it was
not trained to privilege, the compiler injects a dense scale description. The
canonical entry is Listing~\ref{lst:bayati}.

\begin{lstlisting}[caption={The bayati grounding clause, verbatim from the
knowledge base. All 15 mode entries follow this template: spelling,
characteristic degree, ornaments, resting tones, colour, negative guidance.},
label={lst:bayati}]
maqam bayati on D - scale: D E-half-flat F G A Bb C D,
the half-flat second degree (E koron) is the expressive soul
of bayati, ornaments: shimmer on E-half-flat, slides
D->E-half-flat, descending resolution G->F->E-half-flat->D,
resting tones: D (tonic), G (dominant),
color: melancholic, warm, yearning - the most expressive
Arabic maqam, avoid western major or minor tonality,
stay in bayati modal world
\end{lstlisting}

\noindent Two choices matter. The microtonal degree is spelled \emph{phonetically}
(``E-half-flat'', ``E koron'') rather than symbolically, because the model
consumes prose, not notation. And the closing clause supplies explicit negative
guidance---the prose analogue of a repulsion term, intended to keep the latent
trajectory inside the mode~\citep{liu2023promptsurvey,brown2020gpt3}.

\subsection{Prompt Synthesis}
\label{sec:synthesis}

The compiler assembles $p_t$ as an \emph{ordered} concatenation of clauses
joined by commas. Ordering is load-bearing---hard constraints are placed first,
before any evocative description:
\[
\begin{aligned}
p_t = \;&[\text{instrument rule / silence}] \,\Vert\, [\text{no-percussion}] \\
      \,\Vert\, &[\text{specific header}] \,\Vert\, [\text{session stage}]
      \,\Vert\, [\text{maqam line}] \\
      \,\Vert\, &[\text{harmonic context / echo}] \,\Vert\, [\text{role assignment}] \\
      \,\Vert\, &[\text{register}] \,\Vert\, [\text{dynamics}] \,\Vert\,
      [\text{space clause}].
\end{aligned}
\]

\paragraph{Instrument-constraint control} Because the player conducts a subset
of instruments in, the prompt must actively \emph{suppress} the rest---a
prompt-level substitute for the stem-level control the API does not expose. The
leading clause names only the active instruments and issues a hard silence
directive for every inactive one (Listing~\ref{lst:mech}, top). Placing it first
means the model reads the mixing constraint before it ever reads ``slow delta
blues''.

\paragraph{Role layering and echo} Each section carries three role registers, so
the same instrument reads differently comping and soloing. The controller picks
the lead from a per-genre preference list (blues: harmonica $>$ lead guitar $>$
piano) and assigns the remainder their \texttt{supporting} text. In
\textsc{Responding} the lead receives its \texttt{responding} text and the rest
hold; in \textsc{LongIdle} the lead receives the richer
\texttt{solo\_responding} text. The most direct expression of interaction is a
literal echo: the compiler converts \texttt{last\_phrase\_pitches} to note names
and instructs an answer (Listing~\ref{lst:mech}, bottom). The blues path
additionally selects harmonic guidance by a strict four-tier
priority---a chord held right now triggers a hard chord lock; otherwise the
inferred twelve-bar region; otherwise the recent pitch context; otherwise the
raw active notes---so the band always comps on the most reliable signal
available.

\begin{lstlisting}[caption={Two prompt-level control mechanisms that substitute
for control the model API does not expose: instrument suppression in place of
stem-level mixing (top), and a phrase echo that turns transcription into
turn-taking (bottom, blues path).},label={lst:mech}]
INSTRUMENT RULE - this track contains ONLY: <active>.
The following produce absolutely no audio and must not appear:
<Harmonica: NO SOUND, Lead Guitar: NO SOUND, ...>.
Generate nothing for those instruments.
%
soloist just played: A -> C -> E -
their pitch world this session: A -> C -> D -> E -> G -
answer using those same pitches with blues character:
slide into the first note, bend with blues phrasing,
resolve back to A - 'I heard you, here is my reply'
\end{lstlisting}

\paragraph{A compiled prompt} Listing~\ref{lst:example-prompt} shows one
complete melodic prompt as assembled for a concrete situation: genre blues; bass,
piano and harmonica conducted in; state \textsc{Responding}; the player has just
played \texttt{A3\,$\to$\,C4\,$\to$\,E4}. Every mechanism above appears in the
fixed clause order. It also exposes a defect: the residual \texttt{scale:
bayati} clause is the maqam field, which the blues path does not clear, leaking
into a blues prompt. We report it rather than repair the example, because it
illustrates the practical value of a text interface---a compiler bug is legible
as a stray English phrase in the artifact the model receives, with no
instrumentation required.

\begin{lstlisting}[caption={A complete compiled melodic prompt (blues,
\textsc{Responding}, harmonica lead), verbatim.},label={lst:example-prompt}]
INSTRUMENT RULE - this track contains ONLY: Bass, Piano, Harmonica.
The following produce absolutely no audio and must not appear:
Rhythm: NO SOUND, Lead Guitar: NO SOUND. Generate nothing for
those instruments., no drums, no drum kit, no hi-hat, no snare,
no kick drum, Bass, Piano and Harmonica, slow delta blues style
in A minor, A blues hexatonic scale (A  C  D  Eb  E  G) - the b3 (C)
and b5 (Eb) are the expressive blue notes, bends up into C and slides
from Eb to E are the signature blues moves, slow and laid-back blues
feel, session building - finding the groove, getting warmer,
scale: bayati, expressive and authentic character, soloist just
played: A3 -> C4 -> E4 - answer with a direct blues response:
slide into the first note, bend through A3 -> C4 -> E4 with blues
phrasing, add vibrato on the held notes, resolve back to A - it must
sound like 'I heard you, here is my reply', Harmonica steps forward -
plays a short expressive blues lick, wailing bends and slides, raw
blues harp phrasing, intimate and soulful - a direct musical answer
to the soloist - others hold the groove underneath: Bass: electric
bass holds deep root notes, barely moving, quiet resonant foundation
under the soloist; Piano: piano holds quiet sustained chords, soft
and warm, minimal movement - a harmonic cushion behind the soloist,
mid-energy soulful blues answer, groove is finding itself - confident
but not yet at full peak, slow and soulful - play each phrase with
long sustained notes, leave space and silence after each answer
\end{lstlisting}

This single string, regenerated only when $\sigma$ changes and paired with
$\theta_t$, is the entirety of what \sysname{} sends to the generator.

\section{Implementation}
\label{sec:implementation}

The client is a React~19/TypeScript single-page application bundled by Vite; all
audio, input and vision capabilities come from native browser APIs. The backend
is a FastAPI service served by Uvicorn, depending on \texttt{google-genai} for
both Lyria RealTime generation and Gemini vision (optical music recognition in
Learn Mode) and on NumPy and \texttt{soundfile} for audio packing. Every hot-path
message travels over a single persistent WebSocket, keeping per-event overhead
low and connection setup off the latency path.

\paragraph{Streaming engine} The generative model is encapsulated in an
asynchronous \texttt{LyriaEngine} wrapping one live music session against the
\texttt{models/lyria-realtime-exp} endpoint. It exposes a small idempotent
surface: \texttt{update\_controls()} pushes a weighted prompt
$[(\text{text},1.0)]$ and a generation config, and caches both so each is
re-sent only when it actually changes; \texttt{play()} and \texttt{stop()} gate
the stream. A background receive loop packs each incoming $48$\,kHz stereo PCM
chunk into a WAV container, base64-encodes it, and enqueues it on a bounded
queue, dropping the oldest chunk when full---trading a dropped frame under load
for a hard cap on end-to-end latency. Stopping an engine drains the queue so
stale audio never plays after the band has been silenced. The engine treats a
normally-completed receive stream and a \texttt{ConnectionClosed}/%
\texttt{APIError} identically, clearing the session handle so the next control
tick reconnects transparently; transient drops and server-side stream expiry are
therefore invisible to the performer.

\paragraph{Two-engine orchestration} On every inbound message---and on each
$250$\,ms tick---the backend loop recomputes the shared state and then advances
each engine under its own gating logic. The melodic engine sounds only once at
least one section is conducted in and the warmup gate has passed; it stays active
while the soloist plays or within four seconds of the last note, after which it
stops and resets its control signature so a fresh prompt is issued on return. The
percussion engine is armed by a flag raised by gesture or voice and then plays
continuously until stopped; its signature is simply
$(\textsc{bpm},\text{style index})$. A BPM change on either engine forces a
disconnect-and-reconnect, and a session detected as ``triggered but not
playing'' is reset and restarted.

\paragraph{Client-side audio} The browser renders the two streams on two
separate \texttt{AudioContext} instances so they do not contend on one
scheduler. Each chunk is base64-decoded and turned into an \texttt{AudioBuffer}
via \texttt{decodeAudioData}. Playback is scheduled with a running ``next start''
cursor: each buffer starts at \texttt{max(currentTime, nextStart)} and the cursor
advances by the buffer duration minus a negative overlap ($40$\,ms melody,
$20$\,ms percussion) intended to crossfade chunk boundaries and hide decoder edge
artifacts. To bound latency after a pause, each scheduler resets its cursor when
the wall-clock gap since the previous chunk exceeds a threshold ($600$\,ms
melody, $800$\,ms percussion); without the reset a new chunk would be scheduled
at a stale future timestamp, producing an audible silent delay.
Because the model's steady-state cadence is ${\sim}2$\,s, the reset path
rather than the crossfade cursor dominates in steady state
(\S\ref{sec:eval-gating}).

\paragraph{Multimodal control} Four modalities funnel into the same session
state. \emph{MIDI}: note-on, note-off and CC64 sustain are decoded per input
device and forwarded with note, velocity, event type and timestamp.
\emph{Gesture}: a MediaPipe Hands~\citep{mediapipe2019} pipeline
(\texttt{maxNumHands}$=2$, confidence $0.7$) infers per-finger extension from
landmark ordering and recognizes two robust static poses---index-only toggles
percussion, open palm toggles the ensemble. Each must be held for $700$\,ms
before acting, followed by a $2$--$3$\,s cooldown, with hold progress surfaced as
a visual cursor. \emph{Voice}: continuous Web Speech recognition, gated by
confidence $\geq0.62$ and at most eight words, parsed against a rule set
requiring an explicit action word paired with a target; a $1.5$\,s deduplication
window prevents double firing. \emph{Learn Mode}: MusicXML and \texttt{.mxl} are
parsed entirely in-browser; PDF scores are rendered to page images with PDF.js
and sent to Gemini vision, which returns a structured note list, with model
fallbacks tried newest-first.

\paragraph{Constraints that shaped the design} Four properties of the deployed
generator determined non-obvious choices. Lyria cannot change tempo on an open
stream, so BPM changes are implemented as reconnects and the percussion bed is
tempo-locked by design. The API exposes no stem-level mixing, so instrument
control is prompt-level and best-effort. Re-prompting on every note caused
audible thrashing, motivating the fingerprint-based gate. And the model has no
notion of the human's own sound, so keeping the AI behind the soloist is done
with client-side attenuation plus explicit space-enforcing clauses in the prompt.
The recurring theme is that fine-grained control over a general model is
obtained without touching weights: the levers are prompt design,
generation-config scheduling, connection-lifecycle management, and client-side
mixing.

\section{Evaluation}
\label{sec:evaluation}

We evaluate \sysname{} along three lines: (i) instrumented measurement of
the deployed pipeline---latency, throughput, and stream stability
(\S\ref{sec:eval-latency}--\S\ref{sec:eval-gating}); (ii) controlled
ablations that attribute output behaviour to specific design choices over
input-identical replayed sessions (\S\ref{sec:eval-ablation}); and (iii) a
perceptual evaluation in which two expert musicians played the deployed
system (\S\ref{sec:eval-pilot}). The pipeline measurements come from a live
blues session---latency, gating, and stability are properties of the
transport and control loop, not of any genre's prompt content---while all
\emph{output-quality} experiments target the system's design centre, Arabic
maqam.

\paragraph{Apparatus and method} The setup mirrors the deployment of
\S\ref{sec:implementation}: a MIDI keyboard driving the front end in Chromium,
the FastAPI backend on the same host, and Lyria RealTime~\citep{lyria2024} over
the public API. We instrumented every pipeline boundary---key capture, WebSocket
send, backend receive, compile completion, prompt push, chunk arrival, client
decode, scheduled start---as a timestamped JSONL timeline. Because both processes
share a host, client and server wall clocks are directly comparable across the
boundary; intra-process durations use the monotonic clock. Per-stage figures are
differences between adjacent timestamps. Because a real-time system must be
judged by its worst cases as well as its typical case, each distribution is
summarized by its median together with its 95th and 99th percentiles (p95 and
p99)---the values below which $95\%$ and $99\%$ of observations fall,
computed by the nearest-rank method. The end-to-end
figure is measured directly rather than summed: when a re-prompt fires, the
backend attaches the client wall-clock timestamp of the causal key press to every
chunk generated under the new prompt, and the client computes the delay from that
press to the moment the first such chunk is scheduled to become audible. We
report a primary session of $8.5$\,min of continuous playing ($1{,}516$ note
events, $758$ onsets, both engines active, $352$ chunks delivered), consistent
with a shorter preliminary session whose figures agree within noise.

\subsection{Latency and Throughput}
\label{sec:eval-latency}

\begin{figure}[tb]
\centering
\includegraphics[width=\linewidth]{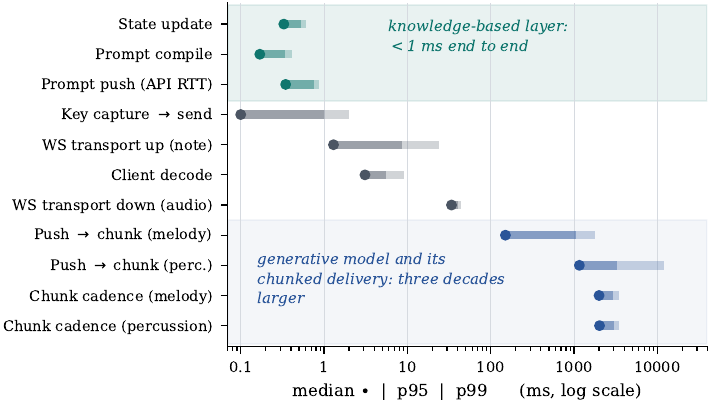}
\caption{Cost of every pipeline stage over the primary $8.5$\,min session: dot
is the median, the solid bar extends to p95 and the faint bar to p99. The three
knowledge-based stages (upper band) together cost under a millisecond even at
the tail; the generative model and its chunked delivery (lower band) sit three
orders of magnitude higher. Sub-$0.1$\,ms values are drawn at the $0.1$\,ms
axis floor.}
\label{fig:stages}
\end{figure}

\begin{figure}[tb]
\centering
\includegraphics[width=\linewidth]{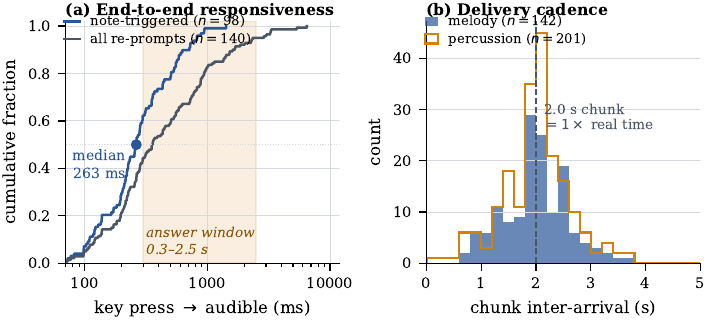}
\caption{(a) Empirical CDF of directly measured key-press-to-audible latency
against the $0.3$--$2.5$\,s answer window the controller itself opens. The
note-triggered median of $263$\,ms lands at the front of the window and the p99
of $945$\,ms well inside it; the heavier ``all re-prompts'' tail includes
heartbeat-fired re-prompts, whose elapsed time from the anchoring key press
includes musical waiting rather than pipeline delay. (b) Chunk inter-arrival
times for both engines, concentrated at the model's $2.0$\,s chunk---generation
runs at $1\times$ real time. Gaps ${\ge}5$\,s, which span playing pauses and
stream restarts rather than steady-state delivery, are excluded ($4$ of $146$
melody and $2$ of $203$ percussion gaps).}
\label{fig:responsiveness}
\end{figure}

Figures~\ref{fig:stages}--\ref{fig:responsiveness} and Table~\ref{tab:latency}
report the measured decomposition. Three observations follow.

First, \emph{the knowledge-based layer is essentially free}. State update plus
prompt compilation totals roughly half a millisecond even at p99, and the push
API round trip is sub-millisecond (Fig.~\ref{fig:stages}); the entire budget is
spent in the generative model and its chunked delivery, which emits uniform
$2.0$\,s chunks at a ${\sim}2.0$\,s cadence
(Fig.~\ref{fig:responsiveness}b). This cleanly separates the
contribution's cost from the backend model's: a faster generator would improve
responsiveness without touching the compiler, and the compiler could run
comfortably inside a far tighter budget than the one it currently sits in.

Second, the direct end-to-end median of $263$\,ms for note-triggered re-prompts
sits at the front of the $0.3$--$2.5$\,s answer window encoded in the session
logic. This is the sense in which turn-taking buys latency headroom: an
accompanist's reply is musically \emph{expected} to arrive somewhat after the
soloist stops, so delay that would be unacceptable for beat-synchronous playing
is perceptually absorbed by the response gap. Three implementation choices bound
the tail deliberately---the bounded queue that drops the oldest chunk rather than
accumulating backlog, the negative scheduling overlap, and the re-prompt gate
that removes per-note compile and API round trips from the hot path. We stress
that fitting the \emph{reaction} budget does not produce beat-level entrainment,
a distinction the expert playing sessions draw sharply (\S\ref{sec:eval-pilot}).

Third, the percussion engine's much larger push-to-chunk figure
(median $1{,}160$\,ms, p99 ${\approx}12$\,s) reflects its lifecycle rather than
its steady state: percussion pushes cluster at style changes and BPM-change
reconnects, where the stream must be re-established before audio resumes.

\begin{table}[tbp]
  \centering
  \caption{Measured latency decomposition, primary session ($8.5$\,min,
  $1{,}516$ note events). All values in ms, nearest-rank percentiles. The
  ``push$\to$next chunk'' rows report arrival of the first chunk after the new
  prompt---a lower bound on when audibly affected audio arrives, since chunk
  content is not attributable to a prompt without audio analysis. The
  chunk-cadence rows are inter-arrival gaps between an engine's consecutive
  server-side emission records ($147$ melody, $204$ percussion;
  Table~\ref{tab:stability}), excluding the six gaps ${\ge}5$\,s that span
  playing pauses and stream restarts rather than steady-state delivery ($4$
  melody, $2$ percussion), hence $n{=}142$ of $146$ and $n{=}201$ of $203$
  gaps. The percussion engine is soloist-decoupled by design and so has no
  end-to-end row.}
  \label{tab:latency}
  \footnotesize
  \setlength{\tabcolsep}{4pt}
  \begin{tabular}{@{}lrrrr@{}}
    \toprule
    \textbf{Stage} & \textbf{median} & \textbf{p95} & \textbf{p99} & $n$ \\
    \midrule
    \multicolumn{5}{@{}l}{\emph{Knowledge-based layer}}\\
    Backend state update                     & $0.3$ & $0.5$ & $0.6$ & $3{,}555$ \\
    Prompt compile                           & $0.2$ & $0.3$ & $0.4$ & $3{,}555$ \\
    Re-prompt push (API round trip)          & $0.3$ & $0.8$ & $0.9$ & $1{,}401$ \\
    \midrule
    \multicolumn{5}{@{}l}{\emph{Transport and client}}\\
    Key capture $\rightarrow$ WS send        & $<\!1$ & $1$ & $2$ & $1{,}516$ \\
    WS transport up (note events)            & $1.3$ & $8.7$ & $24.2$ & $1{,}516$ \\
    WS transport down (chunks)               & $33.9$ & $40.8$ & $44.1$ & $352$ \\
    Client decode $+$ schedule               & $3.1$ & $5.5$ & $9.0$ & $352$ \\
    \midrule
    \multicolumn{5}{@{}l}{\emph{Generative model}}\\
    Push $\rightarrow$ next chunk (melody)   & $150$ & $1{,}070$ & $1{,}799$ & $140$ \\
    Push $\rightarrow$ next chunk (perc.)    & $1{,}160$ & $3{,}250$ & $11{,}934$ & $30$ \\
    Chunk cadence (melody)                   & $1{,}999$ & $2{,}978$ & $3{,}459$ & $142$ \\
    Chunk cadence (percussion)               & $2{,}024$ & $2{,}981$ & $3{,}482$ & $201$ \\
    \midrule
    \multicolumn{5}{@{}l}{\emph{End-to-end (directly measured)}}\\
    Note-triggered re-prompts                & $263$ & $818$ & $945$ & $98$ \\
    All re-prompts                           & $358$ & $2{,}334$ & $5{,}355$ & $140$ \\
    \bottomrule
  \end{tabular}
\end{table}

\subsection{Gating and Stream Stability}
\label{sec:eval-gating}

\begin{figure}[tb]
\centering
\includegraphics[width=\linewidth]{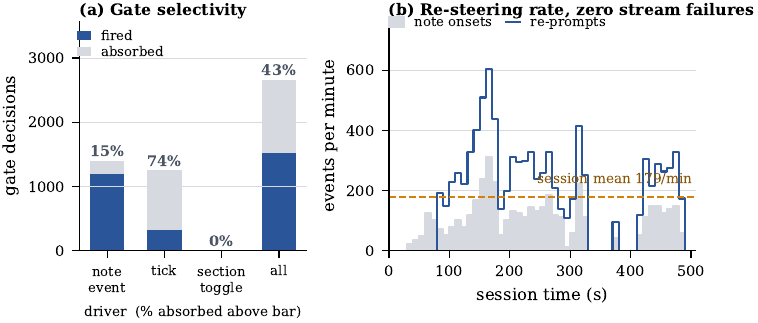}
\caption{(a) Gate decisions by driver, split into re-prompts fired and decisions
absorbed, with the absorbed percentage above each bar: the gate absorbs $74\%$
of heartbeat-driven decisions and $15\%$ of note-driven ones. (b)
Re-steering rate against note-onset rate in $10$\,s bins: re-prompts track
playing activity closely and average $179$ per minute, with zero queue drops,
reconnects, or stream failures across the session.}
\label{fig:gating}
\end{figure}

The central result of this subsection is one of \emph{stability}: the
pipeline tolerates an order of magnitude more re-steering than its design
assumed. Over the primary session, $1{,}517$ gate fires ($179$ per
minute---roughly one re-steer per generated $2$\,s chunk) reached the engines
as $1{,}401$ API pushes ($166$ per minute), and the stream exhibited zero
failures: no reconnects, no stream terminations, and no bounded-queue drops,
with delivery continuing at the model's native cadence throughout
(Fig.~\ref{fig:gating}b, Table~\ref{tab:stability}). The practical ceiling on
re-prompt frequency is therefore set by musical coherence---how often the
model \emph{should} be re-steered---rather than by transport or model
stability. We scope the claim precisely: two service-side disconnects
(\texttt{1011 service unavailable}) occurred in the same log, at $631$\,s and
$684$\,s---after playing had stopped, during an idle tail in which the
browser was left connected. Both were handled by the engine's self-healing
path; they bear on idle-session lifetime, not on stability under load.

The gate itself absorbed $43.1\%$ of the $2{,}665$ decisions it evaluated:
$74.4\%$ of heartbeat-driven decisions and $15.1\%$ of note-driven ones
(Fig.~\ref{fig:gating}a). The lower note-driven selectivity traces to the
harmony fingerprint, whose $2$-semitone pitch-class buckets over a $2$\,s
window are fine enough that most new melodic pitches change the bucket set
during active playing. The fingerprint's granularity is a tunable parameter,
and coarsening it---or adding hysteresis---is a directly measurable
refinement (\S\ref{sec:future-work}). On the client side, chunk-boundary
continuity in steady state rests chiefly on the regularity of chunk arrival
(Fig.~\ref{fig:responsiveness}b); the crossfade cursor of
\S\ref{sec:implementation} engages mainly during burst deliveries, and
re-anchoring its reset condition to the scheduled playback horizon rather
than the arrival gap is a mechanical refinement.

\begin{table}[tbp]
\centering
\caption{Measured gating and streaming-stability summary, primary session.
Failure counters are for the $8.5$\,min active window; the two service-side
disconnects in the same log occurred later, in the idle tail
(\S\ref{sec:eval-gating}). The per-engine chunk split is taken from client-side
delivery records; the server-side emission log captured $351$ of the $352$
deliveries ($147$ melody, $204$ percussion), one melody log write having been
dropped at a stream boundary.}
\label{tab:stability}
\footnotesize
\setlength{\tabcolsep}{5pt}
\begin{tabular}{@{}p{0.55\columnwidth}r@{}}
\toprule
\textbf{Quantity} & \textbf{Value} \\
\midrule
Session length / note events / onsets & $8.5$\,min / $1{,}516$ / $758$ \\
Gate decisions evaluated & $2{,}665$ \\
\quad absorbed, all drivers & $1{,}148$ ($43.1\%$) \\
\quad absorbed, note-driven only & $212$ of $1{,}402$ ($15.1\%$) \\
\quad absorbed, heartbeat-driven & $936$ of $1{,}258$ ($74.4\%$) \\
Gate fires / API pushes & $1{,}517$ / $1{,}401$ \\
Re-prompt rate (fires; pushes) & $179$; $166$ min$^{-1}$ \\
Audio chunks delivered (melody; perc.) & $352$ ($148$; $204$) \\
Bounded-queue drops & $0$ \\
Lyria reconnects; stream terminations & $0$; $0$ \\
\bottomrule
\end{tabular}
\end{table}

\subsection{Perceptual Evaluation: Expert Playing Sessions}
\label{sec:eval-pilot}

Because several of the system's target properties---coherent dialogue,
idiomatic ensemble character, timing fit---are ultimately perceptual, we
complemented the instrumented measurements with playing sessions in which two
expert musicians independently played the deployed system and provided
qualitative feedback. We characterize the scope precisely: these sessions
constitute a small-scale perceptual evaluation by domain experts interacting
with the live system, not a controlled listening study with a larger
participant pool; the latter remains future work (\S\ref{sec:future-work}).

The most consistent feedback concerned \emph{temporal} responsiveness. The
performers perceived the accompaniment as holding an essentially constant
tempo rather than breathing with their playing---the ensemble kept its own
clock. This observation is consistent with the system's construction. The
generator's tempo is fixed for the lifetime of a stream; the percussion bed
is tempo-locked by design; and the melodic engine's ``following'' is
\emph{semantic} rather than \emph{metric}---the compiler tracks register,
density, dynamics, harmony, and phrase boundaries, but performs no beat
tracking on the performer's onsets, so nothing in the control loop re-anchors
generation to the actual pulse. The compiler currently controls \emph{what}
the accompaniment plays more precisely than \emph{when}.

This finding has three consequences. It identifies \emph{tempo and timing
fit} as a rated construct---distinct from the reaction-latency sense of
responsiveness measured in \S\ref{sec:eval-latency}---for the planned
larger-scale study. It sharpens the latency argument: turn-taking headroom
absorbs reaction delay but does not substitute for beat-level entrainment.
And it places beat tracking with tempo re-anchoring at musically safe
boundaries at the head of the future-work agenda.

\subsection{Objective Ablation Results}
\label{sec:eval-ablation}

Three ablations are measurable without listeners, and we report them here as
executed experiments. The remaining, inherently perceptual comparisons are
deferred to the planned controlled study (\S\ref{sec:future-work}).

\paragraph{Method: input-identical replay} All conditions are driven by a
\emph{replay harness} that feeds the recorded input stream of the primary
session---every note event at its original pacing---through the deployed
backend over the same WebSocket path a live client uses. Conditions therefore
receive byte-identical performer input and differ only in a single
environment-selected flag: condition~B replaces the maqam grounding clause
with a generic Western-scale phrase of matched length; condition~C drops the
leading instrument-constraint (silence) clause; the gating arms set the
re-prompt gate to \emph{on}, \emph{off} (fire on every evaluation), or
\emph{static} (compile once, never re-steer). Every session log records its
own condition flags in its first record, so each run is self-describing, and
every generated audio chunk is written to disk tagged with the prompt
sequence number under which it was generated. Because the generator is a live
stochastic service, identical inputs do not produce identical audio; we
therefore replay multiple arms per condition and report distributions.
Replay scripts, per-arm logs, and analysis code are released with the paper.

\begin{figure}[tb]
\centering
\includegraphics[width=\linewidth]{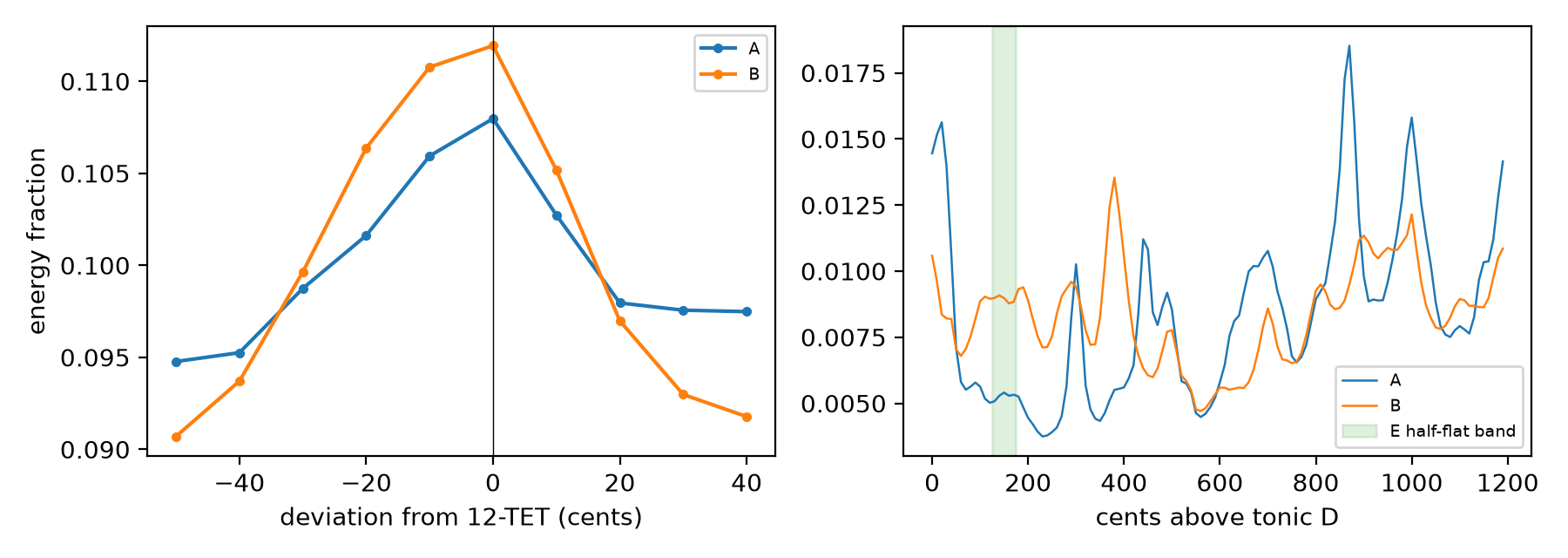}
\caption{Maqam-grounding ablation, three input-identical arms per condition
pooled. (Left) Constant-$Q$ energy versus deviation from the nearest 12-TET
semitone. (Right) Pitch-class energy in cents above the tonic D; the shaded
band marks the half-flat second degree ($125$--$175$ cents) that maqam bayati
prescribes. Aggregate energy shows no concentrated quarter-tone peak, while
frame-level statistics (text) show the grounded lead line visiting the
half-flat band significantly more often.}
\label{fig:microtonal}
\end{figure}

\paragraph{Maqam grounding (A vs.\ B)} Three arms per condition
($553$ vs.\ $521$ generated melody chunks; ${\sim}18$\,min of audio each)
under the bayati configuration, analyzed two ways: a polyphony-safe
constant-$Q$ energy profile at $10$-cent resolution, and pYIN predominant-$F_0$
frame statistics (Fig.~\ref{fig:microtonal}). The grounding clause
demonstrably shapes pitch content, though not deterministically: with
grounding, the tonic D is the most energetic pitch class pooled (share
$0.140$) and in two of three arms ($0.175$, $0.163$), with one arm failing
to anchor ($0.080$, ranked sixth); without grounding D never ranks first in
any arm (pooled $0.094$, ranked fourth). The per-arm spread is itself a
finding about steering a stochastic hosted generator: grounding shifts the
distribution of renderings rather than guaranteeing any single one. By frame statistics, $22.4\%$ of grounded voiced frames lie
${\ge}35$ cents off the 12-TET grid versus $14.9\%$ ablated
($95\%$ CIs $[21.7,23.2]$ vs.\ $[14.0,15.7]$), and---most directly---when
the melodic line visits the second-degree region above D, $59.4\%$ of
grounded frames fall in the half-flat band ($125$--$175$ cents, E~koron)
against $24.1\%$ ablated ($n{=}650$ vs.\ $1{,}530$ frames). At the same time
the \emph{aggregate} energy profile shows no concentrated $150$-cent peak:
second-degree energy splits nearly evenly across E$\flat$/E-half-flat/E
in both conditions, a result robust to harmonic--percussive separation. In
summary, prompt-level grounding elicits genuinely microtonal \emph{melodic
inflection}---the lead line spends significantly more time at quarter-tone
pitch---while rendering the half-flat second as a stably tuned scale degree
remains beyond prompt-level control alone.

\paragraph{Instrument suppression (A vs.\ C)} An audit with the PANNs Cnn14
AudioSet tagger (clip-probability threshold $0.1$, $95\%$ Wilson intervals,
same-day arms to control for service drift) finds that the leading
instrument-constraint clause does not measurably suppress the excluded
section instruments; detections of the excluded cello class were, if
anything, more frequent with the clause present, consistent with the known
tendency of negative prompting to evoke what it names. Absolute rates carry
a caveat---AudioSet has no oud class, so plucked-string content may register
as cello in both conditions. Prompt-level prohibition is therefore not a
substitute for stem-level API control, and we bound the corresponding claim
accordingly (\S\ref{sec:limitations}).

\paragraph{Gating ablation (on / off / static)} Input-identical replays under
the three gate modes (Table~\ref{tab:ablation}). Disabling the gate raises
the push rate from $173$ to $202$ pushes/min---every gate evaluation becomes
an API push---yet the stream again exhibits \emph{zero} failures: no queue
drops, no reconnects, no terminations, and an unchanged $2.0$\,s delivery
cadence. This replicates, under deliberately adversarial re-steering, the
stability result of \S\ref{sec:eval-gating}: the gate's measurable
contribution is API economy ($43.9\%$ of decisions absorbed; $16\%$ fewer
pushes), not stream protection. The static arm isolates the compiler's
value: audio continues to arrive at the model's native cadence, but
trigger-to-chunk latency collapses from a $470$\,ms median to
${\sim}35$\,s---the stream simply stops tracking the performer. The replayed
gate-on arm also reproduces the live session's gating profile ($43.9\%$
vs.\ $43.1\%$ absorbed), validating the replay methodology itself.

\begin{table}[tbp]
  \centering
  \caption{Gating ablation over input-identical replays of the primary
  session. ``e2e'' is trigger$\to$chunk-received, reconstructed by the replay
  client (${\sim}3$--$9$\,ms tighter than the live browser measurement); the
  static arm's e2e reflects chunks no longer being attributable to recent
  input rather than pipeline delay.}
  \label{tab:ablation}
  \footnotesize
  \setlength{\tabcolsep}{4pt}
  \begin{tabular}{@{}lrrr@{}}
    \toprule
    & \textbf{on} & \textbf{off} & \textbf{static} \\
    \midrule
    Gate decisions            & $2{,}870$ & $2{,}946$ & $2{,}946$ \\
    \quad fired               & $1{,}611$ & $2{,}946$ & $10$ \\
    \quad absorbed            & $43.9\%$  & $0\%$     & $99.7\%$ \\
    API pushes (min$^{-1}$)   & $173$     & $202$     & $1.1$ \\
    Queue drops / reconnects / stream ends & $0$ & $0$ & $0$ \\
    Chunk cadence median (ms) & $2{,}006$ & $1{,}999$ & $2{,}005$ \\
    e2e median (ms)           & $470$     & $457$     & $34{,}828$ \\
    \bottomrule
  \end{tabular}
\end{table}

\section{Discussion}
\label{sec:discussion}

\subsection{Why Prompt Compilation Works}
\label{sec:why-prompt-compilation}

Interposing a natural-language prompt between the estimated performance state and
the generator---rather than exposing the model to raw MIDI, learned latents, or a
bespoke control API---buys four properties at once.

It is a \emph{model-agnostic} surface: every current text-conditioned music
model~\citep{lyria2024,copet2023musicgen,magentart2024} accepts free-form prose,
so compiling state into prose rather than into one model's private
parameterization treats the generator as an interchangeable backend, and
retargeting is a matter of re-tuning phrasing and the parameter table rather than
re-architecting the pipeline. It requires \emph{no training}: no paired
soloist/accompaniment corpus, no gradient step, no adapter---all the musical
knowledge the system contributes (Table~\ref{tab:kb}) is deterministic,
human-authored prose, which lets a small team encode substantial expertise
without data-collection or training infrastructure. It is \emph{inspectable and
editable} clause by clause: every accompaniment decision is traceable to a
specific branch of the compiler, a musician can read the exact string sent to the
model and edit the phrasing to change the behaviour, and---as the stray
\texttt{scale: bayati} clause in Listing~\ref{lst:example-prompt}
shows---compiler defects surface as legible English rather than as a silent shift
in a latent. And it is \emph{deterministic}, which is what makes gating
sound at all: because the compiler is a pure function of state, the system can
cheaply detect whether a re-prompt would be musically meaningful before paying
for it.

The measurements add a fifth, quantitative property. The knowledge-based layer
costs under a millisecond end to end (Fig.~\ref{fig:stages}) against a
generator budget three decades larger. Symbolic control of a streaming
foundation model is therefore not a performance compromise: on this pipeline the
expert system is free, and the design question is entirely about what the prose
should say and how often it should change.

In this respect \sysname{} stands in the lineage of knowledge-based systems for
music such as CHORAL~\citep{ebcioglu1988choral}---expertise captured in an
explicit, human-readable representation and applied by deterministic rules---with
the effector replaced. That substitution is the transferable idea. The
architecture it embodies (estimate live human state $\to$ compile to prose $\to$
steer a streaming generator $\to$ render $\to$ loop) is not specific to music: the
same loop could plausibly drive live visuals from pose estimation, steer a text
generator from a running model of a document and its recent edits, or extend code
from the evolving structure of a program. In each case the load-bearing
engineering is identical to ours---a deterministic, domain-informed compiler plus
a policy for when to re-steer. We do not claim to have validated these transfers;
we note only that the pattern appears reusable.

\subsection{Cultural and Microtonal Fidelity}
\label{sec:cultural-fidelity}

Prompt-level grounding is a lightweight and inspectable path toward non-Western
idioms~\citep{shahriar2021maqam,ap2016makam}: it needs no specialized model and no
retraining, and the entire cultural encoding is prose that a maqam expert can
read, critique, and correct. But a lightweight instructional path is not faithful
rendering, and the distinction must be drawn sharply. The prompt can
\emph{instruct} the model to shimmer on a half-flat E and to slide
$D\!\rightarrow\!E\text{-half-flat}$; whether the model \emph{renders} that degree
at its true pitch is a separate question governed by its training data. General
text-to-music models are trained predominantly on twelve-tone equal-tempered
Western material~\citep{agostinelli2023musiclm,copet2023musicgen}, and their
internal representation of pitch may under-represent quarter-tone intervals
entirely. The pitch analysis of \S\ref{sec:eval-ablation} now gives this
question a measured, partial answer. Grounding does not merely decorate: it
anchors generation on the prescribed tonic and moves a significantly larger
share of the melodic line off the equal-tempered grid, with the second degree
visited predominantly in the half-flat band. But the aggregate tuning
evidence is equally clear that the half-flat second is not rendered as a
\emph{stable} scale degree---the model inflects toward the quarter-tone
rather than tuning to it. Whether that inflection-without-anchoring profile
is heard by maqam-trained musicians as expressive authenticity or as
out-of-tune approximation is a perceptual question that pitch measurement
alone cannot settle; it is a central hypothesis for the controlled listening
study planned as future work (\S\ref{sec:future-work}).

\subsection{Limitations}
\label{sec:limitations}

We describe \sysname{} as a working system and a methodological contribution, not
an empirically validated result. Five limitations bound the claims.

\emph{No large-scale user study.} We report instrumented system
measurements, objective output-quality ablations, and a qualitative
perceptual evaluation with two expert musicians (\S\ref{sec:eval-pilot});
a controlled listening study with a larger participant pool remains to be
run. The live measurement session was performed by the first author, and the
ablation arms replay that same performance, so behavioural coverage across
players and genres remains to be established.

\emph{Constraint language is not honoured.} Because the API exposes no
stem-level control, unwanted instruments are addressed purely through negative
prompt language---and the audit of \S\ref{sec:eval-ablation} shows this
mechanism failing: excluded-instrument detections were more frequent
\emph{with} the silence clause than without it. More broadly the whole
control scheme assumes the generator obeys constraint-style instructions
(avoid these tonalities, exclude vocals, hold this tempo), and when it does
not, the compiler has no lower-level fallback; per-stem API control, where it
becomes available, is the principled fix.

\emph{Dependence on a proprietary hosted model, and a latency floor set by it.}
Lyria RealTime is closed and subject to availability, quota and behavioural
change outside our control; its inability to re-tempo a live stream shapes the
architecture, and its $2$\,s chunk granularity sets the floor beneath the
measured $263$\,ms median. Two service-side disconnects in our own log
(\S\ref{sec:eval-gating}) illustrate the exposure.

\emph{Narrow, hand-authored coverage in places.} Harmonic inference is rule-based
and partly blues-specific---the implied-chord logic infers a twelve-bar A7/D7/E7
region from a pitch-class histogram and does not generalize to arbitrary keys
without further rules. The state estimator models one player on one keyboard and
has no notion of multiple simultaneous performers. Voice commands are
English-only, a notable gap given the system's Arabic-music focus. And gesture
control depends on lighting, camera placement and a hold-plus-cooldown protocol,
and has not been evaluated under stage conditions.

\subsection{Future Work}
\label{sec:future-work}

Four directions follow directly from the measurements. \emph{Beat tracking and
tempo following} is the priority: estimate the performer's tempo and beat phase
from the onset stream---the estimator already computes notes-per-second and
inter-onset gaps---and re-anchor both engines to it, absorbing the reconnect cost
at musically safe boundaries such as phrase ends and long idles.
\emph{Recalibrating the gate} is now a measurable exercise rather than a guess:
coarsen the harmony fingerprint or add hysteresis, and sweep re-prompt frequency
against perceived coherence---the gating ablation establishes that stability is
not the binding constraint, so the sweep is bounded only by musical coherence.
\emph{Perceptual validation} means scaling the expert playing sessions of
\S\ref{sec:eval-pilot} into a controlled listening study, whose hypotheses
the executed ablations have sharpened;
the replay harness additionally enables objective MIR measures not yet
computed, such as harmonic agreement with the soloist and rhythmic alignment.
And \emph{reducing platform dependence} means
retargeting the compiler to an open-weights streaming
model~\citep{magentart2024} for on-device operation, which would remove both the
network latency floor and the proprietary-API exposure. Further out, a hybrid
compiler could tune phrasing or timing from performance data while retaining the
inspectable text interface, the estimator could track several simultaneous
soloists, and per-stem control---if future APIs expose it---would replace
best-effort prompt suppression with guaranteed mixing.

\section{Conclusion}
\label{sec:conclusion}

We presented \sysname{}, a real-time accompaniment system built around one
methodological idea: a deterministic, musically-informed compiler from live
performance state to natural-language prompts. Architecturally it is a
knowledge-based system in the classical sense---an inspectable hand-authored
knowledge base, a working memory holding the estimated performance state, and a
rule-based four-state inference layer---whose actuator is prose consumed by a
pretrained streaming generator rather than a symbolic effector. That substitution
is what lets the system obtain idiomatic, microtonal call-and-response
accompaniment for Arabic maqam performance with no fine-tuning of any
kind---and to retarget to eight further Western genres at the cost of
authoring prompt fragments.

Instrumenting the deployed pipeline sharpens the claim in both directions. The
knowledge-based layer costs under a millisecond against a generator budget three
decades larger, so symbolic control of a streaming foundation model is
essentially free; the median key-to-audible-update latency of $263$\,ms lands at
the front of the answer window the controller itself opens; and the stream
tolerates $179$ re-prompts per minute without a single failure---and, in the
gating ablation, $202$ pushes per minute with the gate disabled entirely.
Controlled ablations over input-identical replayed sessions attribute
behaviour to design choices: maqam grounding shifts generation toward the
tonic and yields significantly more quarter-tone melodic content; a static
prompt lets trigger-to-audio latency collapse to tens of seconds, quantifying
what the compiler buys. The same instrumentation also marks the current
limits of prompt-level control---most notably, constraint-style prohibition
is no substitute for stem-level mixing. A perceptual evaluation with two
expert musicians localizes the remaining gap precisely: the compiler controls
\emph{what} the accompaniment plays far more accurately than \emph{when} it
plays it.

The broader significance is that much of the intelligence a responsive creative
partner needs can be located not in model weights but in a transparent,
knowledge-based compiler that translates live human state into language a general
model already understands---and that on current hardware this layer is cheap
enough to be free. As streaming generative models improve, we expect the value of
such lightweight, inspectable, idiom-aware control layers to grow.

\section*{CRediT authorship contribution statement}
\textbf{Jiaxin Du:} Conceptualization, Methodology, Software, Investigation,
Writing -- original draft, Writing -- review \& editing.
\textbf{Yong Zhuang:} Validation, Investigation, Writing -- review \& editing.
\textbf{Haoyu Li:} Investigation, Writing -- review \& editing.

\section*{Declaration of competing interest}
The authors declare no known competing financial interests or personal
relationships that could have appeared to influence the work reported in this
paper.

\section*{Data availability}
The replication package for this article is deposited on figshare
(DOI: \texttt{10.6084/m9.figshare.XXXXXXX}). The deposit contains
the system source code; the primary-session instrumentation log underlying
Figs.~\ref{fig:stages}--\ref{fig:gating} and
Tables~\ref{tab:latency}--\ref{tab:stability}; the replay harness, condition
flags, and per-arm session logs for every ablation arm of
\S\ref{sec:eval-ablation} (each log records its own condition in its first
record); a run manifest mapping arms to logs; the analysis scripts and
their JSON outputs behind Fig.~\ref{fig:microtonal} and
Table~\ref{tab:ablation}; and the generated audio chunks
(${\sim}420$\,MB) for every reported experiment arm. Replaying an arm
reproduces the input exactly; because the deployed generative model (Google
Lyria RealTime) is a hosted stochastic service accessed through its public
API, regenerated audio varies across runs, and the deposited audio and
analysis outputs document the arms reported here.

\appendix

\section{Genre Catalogue}
\label{app:genres}

Beyond its Arabic-maqam core, the compiler carries eight further Western
genres (Table~\ref{tab:genres}). They exist chiefly as evidence for the
adaptation-cost claim of \S\ref{sec:related}: each was added by authoring a
genre configuration---section list and order, per-role prompt language, a
lead-preference list, header text, generation-parameter rows, and percussion
patterns---with no change to the estimator, controller, or transport. Arabic
and blues use bespoke controllers with idiom-specific harmonic and echo
logic; the remaining genres share a generic controller driven entirely by
the configuration record. Output quality outside the Arabic design centre is
not evaluated in this paper and is expected to be weaker, in proportion to
the depth of each genre's authored knowledge.

\begin{table}[tbp]
\centering
\caption{The nine genres: instrument sections (low$\rightarrow$high), lead
instrument, and controller.}
\label{tab:genres}
\footnotesize
\begin{tabular}{@{}llll@{}}
\toprule
\textbf{Genre} & \textbf{Sections (low$\rightarrow$high)} & \textbf{Lead} & \textbf{Controller} \\
\midrule
vintage arabic & cello, oud, strings, qanun, nay & nay & bespoke (takht) \\
blues          & bass, rhythm gtr, piano, harmonica, lead gtr & harmonica & bespoke \\
jazz fusion    & upright bass, piano, jazz gtr, sax, trumpet & saxophone & generic \\
rock \& roll   & bass, rhythm gtr, piano/keys, lead gtr & lead guitar & generic \\
classical      & cello, viola, violin, oboe & oboe & generic \\
modern pop     & bass synth, pad synth, keys, vocal aah & vocal aah & generic \\
flamenco       & cello, violin, flamenco gtr & flamenco gtr & generic \\
cinematic      & cello, strings, brass pad, piano/harp & piano & generic \\
funk           & bass, rhythm gtr, keys/organ, horns & horn section & generic \\
\bottomrule
\end{tabular}
\end{table}

\bibliographystyle{elsarticle-harv}
\bibliography{references}

\end{document}